\documentclass[aps,pra,letterpaper,reprint,showpacs,nofootinbib,superscriptaddress]{revtex4-1}

\usepackage{tikz}
\usepackage{pgfplots}
\usetikzlibrary{plotmarks}
\usetikzlibrary{arrows,decorations.pathmorphing,backgrounds,positioning,fit,petri}
\tikzset{every text node part/.style={font=\footnotesize},>=latex'} 
\usetikzlibrary{calc} 

\usepackage{mathtools} \mathtoolsset{showonlyrefs}

\usepackage{txfonts}
\usepackage{eucal}
\usepackage{microtype}

\usepackage{hyperref}
\hypersetup{
	bookmarksopen=true,%
	bookmarksnumbered=true,%
	colorlinks=true,%
    	linkcolor=blue,
    	citecolor=blue,
    	filecolor=blue,
    	urlcolor=blue,
	pdfstartview=FitH,%
	pdfnewwindow=true
}

\usepackage{graphicx}

\newcommand{\ket}[1]{\lvert#1\rangle} 
\newcommand{\braopket}[3]{\langle #1 | #2 | #3\rangle} 
\newcommand{\expect}[1]{ \langle #1 \rangle} 
\newcommand{\expectt}[1]{ \langle\!\langle #1 \rangle\!\rangle} 

\makeatletter
\newcommand{\vast}{\bBigg@{2}}
\newcommand{\Vast}{\bBigg@{4}}
\makeatother

\newcommand{\me}{\mathrm{e}}
\newcommand{\mi}{\mathrm{i}}

\newcommand{\dif}{\mathrm{d}}
\usepackage{bm} \renewcommand{\vec}[1]{\bm{#1}}
\newcommand{\abs}[1]{\lvert#1\rvert}

\DeclareMathOperator{\ain}{\mathit{a}_\text{in}}
\DeclareMathOperator{\aind}{\mathit{a}_\text{in}^\dag}
\DeclareMathOperator{\aout}{\mathit{a}_\text{out}}
\DeclareMathOperator{\aoutd}{\mathit{a}_\text{out}^\dag}

\DeclareMathOperator{\aRin}{\mathit{r}_\text{in}}
\DeclareMathOperator{\aRind}{\mathit{r}_\text{in}^\dag}
\DeclareMathOperator{\aLin}{\mathit{\ell}_\text{in}}
\DeclareMathOperator{\aLind}{\mathit{\ell}_\text{in}^\dag}
\DeclareMathOperator{\aRout}{\mathit{r}_\text{out}}
\DeclareMathOperator{\aRoutd}{\mathit{r}_\text{out}^\dag}
\DeclareMathOperator{\aLout}{\mathit{\ell}_\text{out}}
\DeclareMathOperator{\aLoutd}{\mathit{\ell}_\text{out}^\dag}
\DeclareMathOperator{\azin}{\mathit{\mathring{a}}_\text{in}} 
\DeclareMathOperator{\azind}{\mathit{\mathring{a}}_\text{in}^\dag}
\DeclareMathOperator{\azout}{\mathit{\mathring{a}}_\text{out}}
\DeclareMathOperator{\azoutd}{\mathit{\mathring{a}}_\text{out}^\dag}

\DeclareMathOperator{\ainout}{\mathit{a}_\text{in/out}}
\DeclareMathOperator{\azinout}{\mathit{\mathring{a}}_\text{in/out}}
\DeclareMathOperator{\aRinout}{\mathit{r}_\text{in/out}}
\DeclareMathOperator{\aLinout}{\mathit{\ell}_\text{in/out}}

\newcommand{\delt}{{\delta t} } 
\newcommand{\rabi}{\omega_\text{R}} 

\newcommand{\sigmams}{\tilde{\sigma}_-}
\newcommand{\sigmaps}{\tilde{\sigma}_+}
\newcommand{\sigmazs}{\tilde{\sigma}_z}
\newcommand{\Os}{\tilde{\Omega}}

\begin{document}
\title{Resonance fluorescence in a waveguide geometry}
\author{ \c{S}\"ukr\"u Ekin Kocaba\c{s}}
\email{ekocabas@ku.edu.tr}
\affiliation{Department of Electrical  \& Electronics Engineering, Ko\c{c} University, Rumeli Feneri Yolu, 34450 Sar{\i}yer, \.Istanbul, Turkey}
\author{Eden Rephaeli}
\email{edenr@stanford.edu}
\affiliation{Department of Applied Physics, Stanford University, Stanford, CA 94305, USA}
\author{ Shanhui Fan}
\email{shanhui@stanford.edu}
\affiliation{Ginzton Laboratory, Department of Electrical Engineering, Stanford University, Stanford, CA 94305, USA}
\date{January, 2012}

\begin{abstract}
We show how to calculate the first- and second-order statistics of the scattered fields for an arbitrary intensity coherent state light field interacting with a two-level system in a waveguide geometry. Specifically, we calculate the resonance fluorescence from the qubit, using input-output formalism. We derive the transmission and reflection coefficients, and illustrate the bunching and anti-bunching of light that is scattered in the forward and backward directions, respectively. Our results agree with previous calculations on one- and two-photon scattering as well as those that are based on the master equation approach.

\end{abstract}
\maketitle

\section{Introduction}
Interaction between an atom and a laser beam tuned close to one of the atomic resonances leads to light emission from the atom---i.e. resonance fluorescence---with a rich set of spectral and temporal properties. Spectrally, as the laser intensity is increased the emitted light will develop symmetric side lobes around the central excitation frequency  and the resulting spectral shape is called the Mollow triplet \cite{Mollow1969}. Temporally, the light emitted will also show anti-bunching with a second order correlation that has a minimum for zero time delay \cite{Kimble1976}. 

Recent advances in integrated optics \cite{OBrien2009,Benson2011} and superconducting circuits \cite{You2011} make it possible to think about quantum systems connected to each other via waveguides that operate at optical or microwave frequencies. For such waveguide embedded systems, the Mollow triplet was observed in the emission spectra from a single superconducting qubit \cite{Astafiev2010} and correlation measurements were also reported \cite{Bozyigit2011,Silva2010,Abdumalikov2011}. These structures were later shown to work as a switch \cite{Abdumalikov2010} or a router \cite{Hoi2011}. In the optical domain, resonance fluorescence was modeled in photonic bandgap waveguides \cite{Florescu2004} and experimentally investigated in a system where a fiber was coupled to a quantum dot \cite{Davanco2009,Davanco2011a}. 

\begin{figure}
\includegraphics{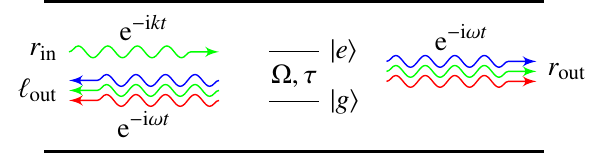}
\caption{(Color online) Schematic of the system under investigation. A right-going coherent state ($\aRin$) at frequency $k$ and an arbitrary intensity, propagating in a waveguide denoted by the long horizontal lines, is incident on a two-level system with energy separation $\Omega$ and a spontaneous emission rate $\tau^{-1}$. After interacting with the qubit, the transmitted ($\aRout$) and the reflected ($\aLout$) light has both a coherent ($\omega=k$) and an incoherent ($\omega\neq k$) component.} \label{fig:Geometry}
\end{figure}

Conventional modeling of resonance fluorescence focuses on light that is emitted in a direction perpendicular to the direction of the laser excitation \cite{Mandel1995,Carmichael2002,Scully2008,Walls2008} which results in anti-bunched statistics. In multi-qubit systems it is possible to observe both bunching and anti-bunching due to the interference of light emission from different qubits \cite{Hennrich2005,Kien2008,Jin2011}. In a waveguide geometry excitation and observation directions are co-linear as shown in Fig \ref{fig:Geometry}. The transmitted amplitudes have contributions from both the incident waves and the emitted waves from the atom. The resonance fluorescence effect is therefore different from the conventional situation. In a previous study based on two-photon scattering off of a qubit embedded in a waveguide, bunching and anti-bunching of light due to the interference of the incoming light with the scattered fields in the transmitted and reflected directions, respectively, was predicted \cite{Shen2007a, *Shen2007}. In this work, as an original contribution, we will extend the two-photon analysis to the case where the excitation is made with an \textit{arbitrary} intensity coherent state.  We will make use of input-output formalism \cite{Gardiner1985} recently generalized to waveguide structures \cite{Fan2010} to derive analytical expressions for the second order correlation functions for the reflected and transmitted fields. We will further show that the low excitation limit of the coherent state solutions agrees very well with the two-photon results. The Mollow triplet will naturally emerge in our analysis. A distinct feature of our analysis is that we can calculate the multi-time correlations without specifically referring to the quantum regression theorem.

The outline of this manuscript is as follows. In Section II we will provide the necessary definitions and derive the single and double time correlations for one-way waveguides by using input-output formalism. In Section III we will extend the analysis to two-way waveguides and derive the spectra of the transmitted and reflected fields. Section IV will have the analysis on double time correlations for the scattered fields where we compare the coherent state and the two-photon results. We will conclude the manuscript in Section V. 

\section{Derivation of the single and double time correlations}
A system consisting of a qubit interacting with photons in a waveguide is described by  the Hamiltonian, $ H = H_0 + H_1$, where \cite{Fan2010}
\begin{align}
\label{eq:Hone}\begin{split}
H_0 &= \int_{-\infty}^\infty \dif\omega \, \omega \, a^\dag_\omega a_\omega, \\
H_1 &= \frac{1}{2} \Omega \sigma_z + \frac{V}{\sqrt{v_g}} \int_{-\infty}^\infty \dif\omega \, \left( \sigma_+ a_\omega  + a^\dag_\omega \sigma_- \right).\end{split}
\end{align}
Here, $\omega$ is the atomic transition frequency, $a^\dag_\omega$ and $a_\omega$ are the creation and annihilation operators for photons at frequency $\omega$, respectively. $\sigma_-$ and $\sigma_+$ are the lowering and raising operators for the qubit, $\sigma_z=[\sigma_+,\sigma_-]$.  $V$ denotes the coupling strength between the atomic states and the waveguide modes, and $v_g$ is the group velocity of the propagating waveguide mode. In the derivation of the Hamiltonian we make the dipole and the rotating wave approximations, linearize the waveguide dispersion around the excitation frequency to obtain the group velocity and assume that the  photons are at a frequency in the vicinity of the excitation wavelength so that the linearization can be justified \cite{Fan2010}. 

We set 
\begin{align}
\frac{1}{\tau} ={}& \pi \frac{V^2}{v_g}, \\
\ain(t) ={}& \frac{1}{\sqrt{2\pi}} \int \dif \omega \, a_\omega(t_0) \me^{-\mi \omega (t-t_0)}, \\
\aout(t) ={}& \frac{1}{\sqrt{2\pi}} \int \dif \omega \, a_\omega(t_1) \me^{-\mi \omega (t-t_1)},
\end{align}
where $\ain$ and $\aout$ are the input and output fields defined long before ($t_0 \rightarrow -\infty$) and long after ($t_1 \rightarrow \infty$) the interaction between the qubit and the the photons takes place. The two fields are related by
\begin{equation}
\aout(t) = \ain(t) - \mi \sqrt{\frac{2}{\tau}} \sigma_-(t). \label{eq:ainout}
\end{equation}
Through the help of the Heisenberg equations of motion and the definitions of the input and output fields, we can write the following set of input-output equations for a single qubit system \cite{Fan2010}
\begin{align}
\frac{\dif \sigma_-(t)}{\dif t} ={} & \mi \sqrt{\frac{2}{\tau}} \sigma_z(t) \ain(t)-\left(\frac{1}{\tau}+\mi \Omega \right)\sigma_-(t), \\
\frac{\dif \sigma_+(t)}{\dif t} ={} & -\mi \sqrt{\frac{2}{\tau}} \aind(t) \sigma_z(t) -\left(\frac{1}{\tau}-\mi \Omega \right)\sigma_+(t) \label{eq:inout},\\ 
\frac{\dif \sigma_z(t)}{\dif t} ={} & -\mi 2 \sqrt{\frac{2}{\tau}} [\sigma_+(t) \ain(t) - \aind(t) \sigma_-(t)] -\frac{2}{\tau} [\sigma_z(t)+1].
\end{align}

In this article we will be interested in the statistics of the scattered fields when a coherent state input is incident on the qubit. We define the incoming coherent state at frequency $k$ as 
\begin{align}
\ket{\alpha_k^+} ={} &\me^{-\abs{\alpha_k}^2/2} \sum_{n=0}^\infty \frac{\alpha_k^n}{\sqrt{n!}} \ket{n_k^+}
= \me^{-\abs{\alpha_k}^2/2} \sum_{n=0}^\infty \frac{\alpha_k^n \aind(k)^n}{n!} \ket{0},
\shortintertext{such that}
\ain(t) \ket{\alpha_k^+} ={} & \frac{1}{\sqrt{2\pi}} \int \dif k' \ain(k') \me^{-\mi k' t} \ket{\alpha_k^+}\\
={} & \frac{\alpha_k}{\sqrt{2\pi}} \me^{-\mi k t} \ket{\alpha_k^+} = \frac{\rabi}{2} \sqrt{\frac{\tau}{2}}\me^{\mi\phi-\mi k t} \ket{\alpha_k^+}. \label{eq:aincoherent}
\end{align}
The value of $\alpha_k$ is in general complex valued. We define $\alpha_k \equiv \abs{\alpha_k} \me^{\mi \phi}$. $\rabi\equiv 2\abs{\alpha_k}/\sqrt{\pi \tau}$ is the Rabi frequency.

The expectation value of an operator, $O$, is given as
\begin{align}
\expect{O} \equiv \braopket{\alpha_k^+}{O}{\alpha_k^+}.
\end{align}
In order to describe resonance fluorescence in a waveguide, three classes of correlation functions will be of importance: ones with one operator, ones with two operators at two different times and ones with three operators at two different times, i.e.
\begin{align}
\vec{c}_1(t=0,t')={}&
\begin{pmatrix}
\expect{\sigma_-(t')} \\
\expect{\sigma_+(t')}\\
\expect{\sigma_z(t')}
\end{pmatrix}, 
\quad \vec{c}_2(t,t')=
\begin{pmatrix}
\expect{\sigma_+(t) \sigma_-(t')} \\
\expect{\sigma_+(t) \sigma_+(t')} \\
\expect{\sigma_+(t) \sigma_z(t')} 
\end{pmatrix},\\
\vec{c}_3(t,t')={}&
\begin{pmatrix}
\expect{\sigma_+(t) \sigma_-(t') \sigma_-(t)} \\
 \expect{\sigma_+(t) \sigma_+(t') \sigma_-(t)}\\
 \expect{\sigma_+(t) \sigma_z(t') \sigma_-(t)} 
\end{pmatrix}. \label{eq:expectSet}
\end{align}
In order to calculate these expectation values, we use input-output equations \eqref{eq:inout} and multiply them from the left and the right with the necessary terms.\footnote{For instance, to get the second set, $\vec{c}_2(t,t')$,  we need to multiply \eqref{eq:inout} evaluated at time $t'$ by $\sigma_+(t)$ from the left.} We then take the expectation values, make use of \eqref{eq:aincoherent} and the commutator $[\ain(t'),\sigma_-(t)]=0$ for $t' \ge t$ \cite{*[{This commutator identity for $t'>t$ is proved using causality in \cite{Gardiner1985}. For $t'=t$, the identity can be directly proved using the definition of the operators, as shown in the appendix of }] [] Rephaeli2011} to arrive at the following set of differential equations for all three classes of expectation values ($n=1,2,3$)
\begin{align}
&\frac{\dif}{\dif t'} \vec{c}_n(t,t') = \vec{B}(t') \vec{c}_n(t,t') + \vec{b}_n \quad \text{where}\\
&\vec{B}=\begin{pmatrix}
-(1/\tau + \mi \Omega) & 0 &  \frac{1}{2}\mi \rabi \me^{-\mi k t'} \me^{\mi \phi} \\
0 & -(1/\tau - \mi \Omega)& -\frac{1}{2}\mi\rabi \me^{\mi k t'} \me^{-\mi \phi}\\
\mi \rabi \me^{\mi k t'} \me^{-\mi \phi} & -\mi \rabi \me^{-\mi k t'} \me^{\mi \phi} & -2/\tau
\end{pmatrix}, \: \vec{b}_n= \begin{pmatrix}
0\\
0\\
b_n
\end{pmatrix}.
\end{align}
These are called the optical Bloch equations with radiative damping. For different $n$, the inhomogeneous term $b_n$ and the initial conditions at $t'=t$ are different: $b_1=-\frac{2}{\tau}$, $b_2=-\frac{2}{\tau}\expect{\sigma_+(t)}$ and $b_3=-\frac{2}{\tau}\expect{\sigma_+(t)\sigma_-(t)}$. Previously, the same results were derived through the help of the quantum regression theorem \cite{Carmichael2002,Walls2008}. However, the derivation here follows naturally within input-output formalism. In Appendix \ref{app:A} we provide the derivation of the general solution to the Bloch equations and in Appendix \ref{app:B} explicit solutions for all $\vec{c}_n$ are listed.

\section{Derivation of the fluorescence spectrum of  the transmitted and the reflected light}
Up till now, our analysis did not distinguish between right- and left-going waves. Indeed, the Hamiltonian we wrote was for a chiral (i.e.~one-way) waveguide. For a regular two-way waveguide where fields propagate in both directions, the Hamiltonian has separate input and output operators for right ($r$) and left ($\ell$) propagating waves \cite{Fan2010}. The equations of motion become
\begin{align}
\frac{\dif \sigma_-}{\dif t} ={} & \mi \sqrt{\frac{2}{\tau}} \sigma_z \aRin+\mi \sqrt{\frac{2}{\tau}} \sigma_z \aLin-\left(\frac{2}{\tau}+\mi \Omega \right)\sigma_-, \\
\frac{\dif \sigma_+}{\dif t} ={} & -\mi \sqrt{\frac{2}{\tau}} \aRind \sigma_z -\mi \sqrt{\frac{2}{\tau}} \aLind \sigma_z -\left(\frac{2}{\tau}-\mi \Omega \right)\sigma_+ \label{eq:inout2},\\ 
\frac{\dif \sigma_z}{\dif t} ={} & -\mi 2 \sqrt{\frac{2}{\tau}} [\sigma_+ (\aRin+\aLin) - (\aRind+\aLind) \sigma_-] -\frac{4}{\tau} [\sigma_z+1].
\end{align}
We can decompose the right and left input/output states as
\begin{align}
\label{decomposition}
\begin{split}
\aRinout(t) &= \frac{\ainout(t)+\azinout(t)}{\sqrt{2}}, \\
\aLinout(t) &= \frac{\ainout(t)-\azinout(t)}{\sqrt{2}},
\end{split}
\end{align}
and as a result arrive at the Hamiltonian, $H=H_0 + H_1$, where
\begin{align}
H_{0}&=\int \dif \omega \, \omega\,( a_\omega^\dag a_\omega + \mathit{\mathring{a}}_\omega^\dag \mathit{\mathring{a}}_\omega), \\
H_{1}& = \frac{1}{2} \Omega \sigma_z + \frac{\sqrt{2}V}{\sqrt{v_g}}\int \dif \omega \, \left( \sigma_+ a_\omega + a_\omega^\dag \sigma_- \right).
\end{align}
The fields $a$ and $\mathit{\mathring{a}}$ are even and odd combinations, respectively, of the right and left propagating fields. The interacting part of the Hamiltonian, $H_1$, depends on $a$ only and the $\mathit{\mathring{a}}$ dependence is solely in the non-interacting part, $H_0$. Except for an additional term in $H_0$,\footnote{Note also that the extra factor of $\sqrt{2}$ in front of $V$ in $H_1$ will lead to a redefinition $\tau\rightarrow\tau'\equiv\tau/2$.} the two-way Hamiltonian is very similar to the chiral Hamiltonian in \eqref{eq:Hone}. Hence, we will be able to make use of the results of the previous section in the analysis of two-way waveguides. To do so, we decompose a right going coherent state with frequency $k$ into two separate---even and odd---channels \cite{Barnett1997}
\begin{align}
&\exp\left[ \alpha \aRind(k) - \alpha^* \aRin(k) \right] \ket{0}\\
&=\exp\left[ \alpha \frac{\aind(k)+\azind(k)}{\sqrt{2}} - \alpha^* \frac{\ain(k)+\azin(k)}{\sqrt{2}}\right] \ket{0} \equiv \ket{\stackrel{\text{(even)}}{\frac{\alpha_k^+}{\sqrt{2}}}; \stackrel{\text{(odd)}}{\frac{\alpha_k^+}{\sqrt{2}}}}\\
\shortintertext{such that}
&\ain(t) \ket{\frac{\alpha_k^+}{\sqrt{2}}; \frac{\alpha_k^+}{\sqrt{2}}} = \azin(t) \ket{\frac{\alpha_k^+}{\sqrt{2}}; \frac{\alpha_k^+}{\sqrt{2}}} = \azout(t) \ket{\frac{\alpha_k^+}{\sqrt{2}}; \frac{\alpha_k^+}{\sqrt{2}}} \label{eq:aintwo}\\
&\qquad= \frac{\alpha_k }{\sqrt{2} \sqrt{2 \pi}} \me^{-\mi k t} \ket{\frac{\alpha_k^+}{\sqrt{2}}; \frac{\alpha_k^+}{\sqrt{2}}} = \frac{\rabi}{2}\sqrt{\frac{\tau'}{2}}\me^{\mi \phi -\mi k t} \ket{\frac{\alpha_k^+}{\sqrt{2}}; \frac{\alpha_k^+}{\sqrt{2}}}, 
\end{align}
where $\tau'\equiv \tau/2$ absorbs the $\sqrt{2}$ factor. As one can see, the odd channel is interaction-free and thus is an eigenstate of $\azin(t)=\azout(t)$ whereas the even channel is subject to $H_1$. Nevertheless, it is the combination of both the even and odd channels that lead to the right- and left-going fields. The two channel expectation value of an operator $O$ is defined as
$$
\expectt{O} \equiv \braopket{\frac{\alpha_k^+}{\sqrt{2}}; \frac{\alpha_k^+}{\sqrt{2}}}{O}{\frac{\alpha_k^+}{\sqrt{2}}; \frac{\alpha_k^+}{\sqrt{2}}}.
$$

In order to calculate the spectral properties of the transmitted fields, we need to calculate the Fourier transform of $\expectt{\aRoutd(t) \aRout(t+\delt)}$ with respect to $\delt$ (see Fig \ref{fig:Geometry}). By using \eqref{decomposition} we can write
\begin{align}
&\expectt{\aRoutd(t) \aRout(t+\delt)}\\
&\quad=\frac{1}{2}\expectt{[\aoutd(t)+\azoutd(t)][\aout(t+\delt)+\azout(t+\delt)]}.\\
\shortintertext{The application of \eqref{eq:ainout} with $\tau\rightarrow\tau'$ results in}
&\quad=\frac{1}{2}\big[ \rabi^2 \frac{\tau'}{2}\me^{-\mi k \delt}-\mi\rabi \me^{-\mi\phi+\mi k t}\expectt{\sigma_-(t+\delt} \\
&\qquad +\mi\rabi \me^{\mi\phi-\mi k (t+\delt)}\expectt{\sigma_+(t)} + \frac{2}{\tau'} \expectt{\sigma_+(t)\sigma_-(t+\delt)} \big].
\end{align}
One can show that the two channel expectation values of operators are the same as their single channel expectation values [i.e. those in \eqref{eq:expectSet}] except for the substitution $\tau\rightarrow\tau'$. The derivation can be made by using \eqref{eq:inout2}, taking the relative expectation values and using \eqref{eq:aintwo} to simplify the results. Therefore, we can use the steady state values from Appendix \ref{app:B} to arrive at
\begin{align}
\expectt{\aRoutd(t) \aRout(t+\delt)}={}&\frac{1}{\tau'} \frac{R^2}{4} \left(\frac{-1+D^2+\frac{1}{2}R^2}{1+D^2+\frac{1}{2}R^2}\right) \me^{-\mi k \delt} \\
&\quad + \frac{1}{\tau'} \expectt{\sigma_+(t)\sigma_-(t+\delt)},
\end{align}
where
\begin{align}
D=(\Omega-k)\tau' \qquad \text{and} \qquad  R= \rabi \tau'.
\end{align}
In order to calculate the Fourier transform of this expression, we need to know $\expectt{\sigma_+(t)\sigma_-(t+\delt)}$ for $\delt<0$ as well.  By using the identity $\expectt{\sigma_+(t+\delt) \sigma_-(t)}=\expectt{\sigma_+(t)\sigma_-(t+\delt)}^*$ we can see that the expectation values for $\delt<0$ are related to those with $\delt>0$ by complex conjugation. The Laplace transform results in Appendix \ref{app:B} thus allow us to calculate the Fourier transform as
\begin{align}
&G^{(1)}_r(\omega) \equiv \mathcal{F}_{\delt}[\expectt{\aRoutd(t) \aRout(t+\delt)} ] =\\ 
&\quad\frac{1}{\tau'} \frac{1}{\sqrt{2\pi}} \frac{\frac{1}{2}R^2}{1+D^2+\frac{1}{2}R^2}
\Bigg[
\pi \delta(\omega-k)\left( D^2+\frac{1}{2}R^2 \frac{D^2+\frac{1}{2}R^2}{1+D^2+\frac{1}{2}R^2}\right)\\
&\qquad+ \frac{R^2}{\tau'^5} \frac{(\omega-k)^2\tau'^2+4+\frac{1}{2}R^2}{\abs{P[-\mi(\omega-k)]}^2}
\Bigg],
\end{align}
where the function $P$ is as defined in \eqref{eqn:ps}. We will use the non-interacting case, that is,
\begin{align}
G^{(1)}_{r_0}\equiv \mathcal{F}_{\delt}[\expectt{\aRind(t) \aRin(t+\delt)} ]= \frac{\sqrt{2\pi} R^2}{4 \tau'}\delta(\omega-k)
\end{align}
for normalization. As a result, the coherent part of the correlation function, one which is proportional to $\delta(\omega-k)$, will be given by
\begin{align}
g^{(1)}_{r_\text{coh}}=   \frac{1}{1+D^2+\frac{1}{2}R^2}\left(D^2+\frac{1}{2}R^2 \frac{D^2+\frac{1}{2}R^2}{1+D^2+\frac{1}{2}R^2}\right). \label{eq:g1R}
\end{align}

\begin{figure}
\includegraphics{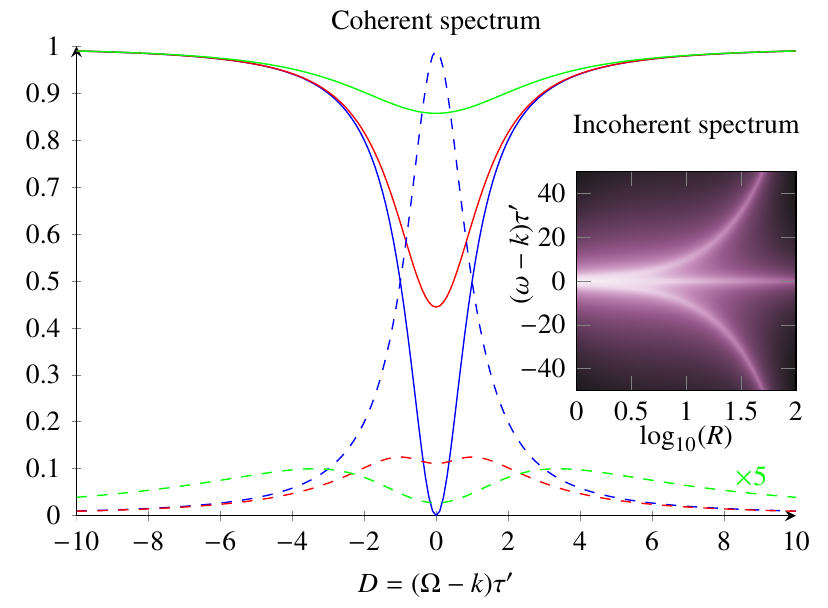}
\caption{(Color online) Coherent part of the transmitted (solid) and reflected (dashed) fluorescence spectrum for $R=\rabi \tau'=\{0.1,2,5\}$ corresponding to the blue, red and green curves respectively. The reflected fluorescence for $R=5$ (dashed green curve) is plotted after being multiplied by five. Inset shows the incoherent part ($\omega \neq k$ case as depicted in Fig \ref{fig:Geometry}) of the spectrum with the Mollow triplet for zero detuning ($D=0$). } \label{fig:Spectrum}
\end{figure}

For reflected fields we need to do a similar analysis for $\expectt{\aLoutd(t) \aLout(t+\delt)}$. By using \eqref{decomposition} and \eqref{eq:ainout} we can see that 
\begin{align}
\expectt{\aLoutd(t) \aLout(t+\delt)}=\frac{1}{\tau'} \expectt{\sigma_+(t) \sigma_-(t+\delt)}.
\end{align}
The Fourier transform of this term is given by
\begin{align}
&G^{(1)}_\ell(\omega)\equiv\mathcal{F}_{\delt}[\expectt{\aLoutd(t) \aLout(t+\delt)}]  \\
&\quad =\frac{1}{\tau'} \frac{1}{\sqrt{2\pi}} \frac{\frac{1}{2}R^2}{1+D^2+\frac{1}{2}R^2} \Big[ \frac{1+D^2}{1+D^2+\frac{1}{2}R^2} \pi \delta(\omega-k) \\
&\qquad + \frac{R^2}{\tau'^5} \frac{(\omega-k)^2\tau'^2+4+\frac{1}{2}R^2}{\abs{P[-\mi(\omega-k)]}^2} \Big].
\end{align}
We again normalize with respect to the non-interacting case, and obtain
\begin{align}
g^{(1)}_{\ell_\text{coh}}=\frac{1+D^2}{\left(1+D^2+\frac{1}{2}R^2\right)^2} \label{eq:g1L}
\end{align}
for the coherently back scattered fields. The incoherent parts of the reflected and transmitted fields are equal to each other and are given by 
\begin{align}
g^{(1)}_{\text{incoh}}=\frac{1}{\pi}  \frac{1}{1+D^2+\frac{1}{2}R^2}  \frac{R^2}{\tau'^5} \frac{(\omega-k)^2\tau'^2+4+\frac{1}{2}R^2}{\abs{P[-\mi(\omega-k)]}^2}.
\end{align}
In Fig \ref{fig:Spectrum} spectral features of the transmitted and reflected fields are plotted. Note that the results \eqref{eq:g1R}--\eqref{eq:g1L} agree with those in \cite{Astafiev2010,Hoi2011}.

\begin{figure*}
\includegraphics{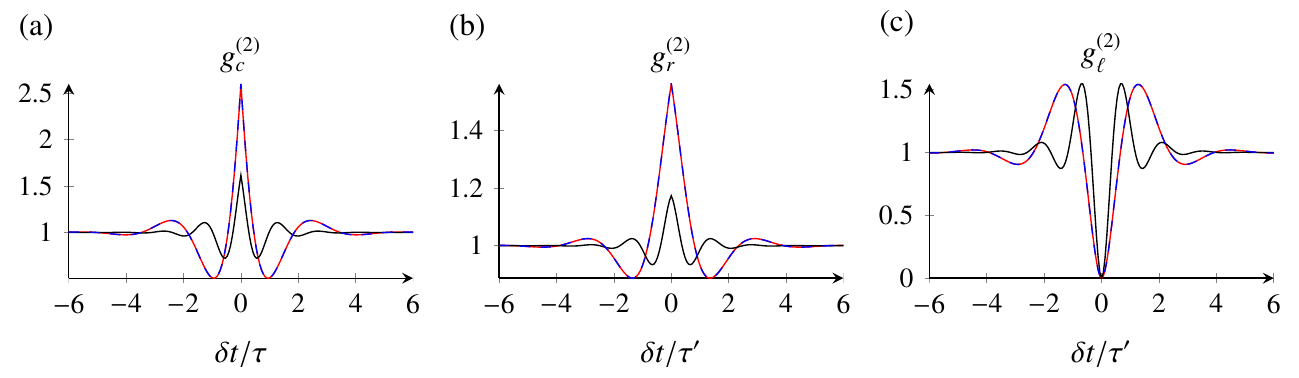}
\caption{(Color online) Plots of $g^{(2)}$ for $D=2$ and $R=\{ 0.2, 4\}$ for the red and black curves, respectively. The dashed blue curve is the normalized two-photon wavefunction. (a) one-mode; (b) two-mode, transmitted; (c) two-mode, reflected case. As can be seen, the two-photon calculations are indistinguishable from resonance fluorescence ones for $R=0.2$ but not for $R=4$. } \label{F:3}
\end{figure*}

\section{Derivation of the second order correlation function of the transmitted and reflected light}
Now that we have calculated various first order correlations and investigated spectral properties of scattered fields, we can start to look into the time dependent statistics of the transmitted and reflected fields. To do so, we will begin by investigating the second order correlation function in a chiral waveguide, $g^{(2)}_c$, given by
\begin{align}
g^{(2)}_c(\delt) = \lim_{t\rightarrow\infty} \frac{\expect{\aoutd(t) \aoutd(t+\delt) \aout(t+\delt) \aout(t)}}{\expect{\aoutd(t)\aout(t)}\expect{\aoutd(t+\delt)\aout(t+\delt)}}.
\end{align}
By using \eqref{eq:ainout} and the results from Appendix \ref{app:B} it can be shown that
\begin{align}
\expect{\aoutd(t)\aout(t)} = \frac{1}{\tau} \frac{R^2}{8}.
\end{align}
Similarly, after some algebra we arrive at the following formula for the Laplace transform of $g^{(2)}_c$
\begin{align}
\mathcal{L}_\delt[g^{(2)}_c(\delt)]=\frac{1}{s}+\frac{8}{1+D^2+\frac{1}{2}R^2} \frac{s \left(s+\frac{1}{\tau}\right)}{P(s)}.
\end{align}
Using the expression above, we can show that
\begin{align}
\lim_{R\rightarrow0}g^{(2)}_c(\delt)&=\frac{\psi^{(2)}_{k,p}(t,t+\delt)}{\frac{1}{\sqrt{2}}[\psi^{(1)}_k(t)\psi^{(1)}_p(t+\delt)+\psi^{(1)}_p(t)\psi^{(1)}_k(t+\delt)]} \\
&= \left| 1 + \frac{4}{(D+\mi)^2} \me^{-\mi\frac{\abs{\delt}}{\tau}( D-\mi)}  \right|^2
\end{align}
where $\psi^{(1)}$ [$\psi^{(2)}$] is the one-photon [two-photon] wavefunction.\footnote{See equations (43) and (120) in \cite{Shen2007} for the one- and two-photon wavefunctions, respectively. Note that the photons are at the same frequency, i.e. $k=p$.} As a result, we have shown that the second order statistics of a low intensity coherent state input and a two-photon input to a qubit are the same.

For the two-mode case, we will need to calculate the correlation functions for the right-going ($r$) and the left-going ($\ell$) fields as
\begin{align}
g^{(2)}_r(\delt) &= \lim_{t\rightarrow\infty} \frac{\expectt{\aRoutd(t) \aRoutd(t+\delt) \aRout(t+\delt) \aRout(t)}}{\expectt{\aRoutd(t)\aRout(t)}\expectt{\aRoutd(t+\delt)\aRout(t+\delt)}}, \\
g^{(2)}_\ell(\delt) &= \lim_{t\rightarrow\infty} \frac{\expectt{\aLoutd(t) \aLoutd(t+\delt) \aLout(t+\delt) \aLout(t)}}{\expectt{\aLoutd(t)\aLout(t)}\expectt{\aLoutd(t+\delt)\aLout(t+\delt)}}.
\end{align}
The normalization terms are given by
\begin{align}
\expectt{\aRoutd(t)\aRout(t)} &= \frac{1}{\tau'} \frac{R^2}{4}\frac{D^2+\frac{1}{2}R^2}{1+D^2+\frac{1}{2}R^2},\\
\expectt{\aLoutd(t)\aLout(t)} &= \frac{1}{\tau'} \frac{\frac{1}{4}R^2}{1+D^2+\frac{1}{2}R^2}.
\end{align}
After some algebra done by the help of an automatic non-commutative algebra system \cite{Supplement} we get
\begin{align}
&\mathcal{L}_\delt[g^{(2)}_r(\delt)]=\\
&\qquad\frac{1}{s}+\frac{1}{A^2 P(s)}\left[ \left( s+\frac{1}{\tau'}\right)\left((1+2A)s+\frac{1+4A}{\tau'}\right)+\frac{D^2}{\tau'^2} \right]
\end{align}
where $A=D^2 + \frac{1}{2}R^2$, and
\begin{align}
\mathcal{L}_\delt[g^{(2)}_\ell(\delt)]={}&\frac{2}{\tau'^2}\left( 1 + D^2 + \frac{1}{2}R^2 \right)\frac{s+\frac{1}{\tau'}}{s P(s)}.
\end{align}
In the limit $R\rightarrow 0$ the second order correlation results for coherent state  and two-photon inputs in a two-mode waveguide can be shown to equal each other, just like in the chiral case (see Fig \ref{F:3}). As was previously predicted, the interference of incoming and scattered fields leads to bunching and anti-bunching in the forward and backward directions, respectively. When $R$ is increased, the response of the qubit gets saturated and there is less bunching in the forward direction but the reflected fields continue to show strong anti-bunching. In \cite{Chang2007,Zheng2010} $g^{(2)}$ was calculated for a low intensity coherent state interacting with a qubit in a waveguide where the qubit was coupled at a rate $\Gamma'$ to non-waveguide modes as well. Our results supplement these previous investigations by analytically describing the scattering of an arbitrary intensity\footnote{We still operate within the bounds of the dipole and the rotating wave approximations used in the derivation of the Hamiltonian.} coherent state off of a qubit for the $\Gamma'=0$ case.

\section{Conclusion}
In this manuscript we used input-output formalism for waveguides to analyze how an arbitrary intensity coherent state scatters off of a qubit embedded in a waveguide. We provided analytical results for the spectra as well as the second-order correlation functions of the transmitted and reflected fields. This work supplements the previous work on two-photon calculations and shows that the two formulations agree for low intensity coherent state inputs. We predicted that the transmitted fields are bunched and the reflected fields are anti-bunched for coherent state inputs, similar to the case for two-photon scattering. Very recent experiments in circuit QED agree with these observations \cite{Hoi2012}. Functional devices---e.g. transistors \cite{Chang2007}, switches \cite{Abdumalikov2010} or routers \cite{Hoi2011}---that make use of multilevel systems require both control signals that are in a coherent state basis and single- or multi-photon Fock states that carry the information. We demonstrated the versatility of input-output formalism with which one can do analysis either based on Fock states to calculate the full scattering matrix, or based on coherent states with an emphasis on correlation measurements. Additionally, it is possible to investigate non-linear effects such as the ac Stark \cite{Schuster2005} and Lamb \cite{Fragner2008} shifts using the methods developed to characterize qubit-coherent state interactions. Lastly, our approach paves the way to calculations involving higher-order correlation functions that become relevant when the qubit is  strongly excited---in a recent cavity QED experiment asymmetry in  time for $g^{(3)}$ was demonstrated \cite{Koch2011}.

\begin{acknowledgments}
This  work is supported by the David \textit{\&} Lucile Packard Foundation. 
\end{acknowledgments}


\appendix

\section{General solution to the Bloch equations} \label{app:A}
In this appendix we will provide the general solution to the differential equation
\begin{align}
\frac{\dif}{\dif t'} \begin{pmatrix}x_1(t,t') \\ x_2(t,t') \\ x_3(t,t') \end{pmatrix} = \vec{B}(t')
\begin{pmatrix}x_1(t,t') \\ x_2(t,t') \\ x_3(t,t') \end{pmatrix}
+
\begin{pmatrix}0 \\ 0 \\ b(t) \end{pmatrix} \quad \text{where} \\
\vec{B}(t')=\begin{pmatrix}
-(1/\tau + \mi \Omega) & 0 &  \frac{1}{2}\mi \rabi \me^{-\mi k t'} \me^{\mi \phi} \\
0 & -(1/\tau - \mi \Omega)& -\frac{1}{2}\mi\rabi \me^{\mi k t'} \me^{-\mi \phi}\\
\mi \rabi \me^{\mi k t'} \me^{-\mi \phi} & -\mi \rabi \me^{-\mi k t'} \me^{\mi \phi} & -2/\tau
\end{pmatrix}
\end{align}
with the initial conditions given at $t'=t$ by $x_1(t,t)$, $x_2(t,t)$ and $x_3(t,t)$. Here $b(t)$ is the inhomogeneous term, independent of $t'$. The solution method we use is the same as the one in \cite{Kimble1976,Mandel1995}. We begin by writing down the equations separately as
\begin{align}
\frac{\dif x_1}{\dif t'} + \left(\frac{1}{\tau} + \mi \Omega\right)x_1 ={}& \frac{1}{2}\mi \rabi \me^{-\mi k t'} \me^{\mi \phi} x_3, \\
\frac{\dif x_2}{\dif t'} + \left(\frac{1}{\tau} -  \mi \Omega\right)x_2 ={}& -\frac{1}{2}\mi \rabi \me^{\mi k t'} \me^{-\mi \phi} x_3, \\
\frac{\dif x_3}{\dif t'} + \frac{2}{\tau} x_3 ={}& \mi \rabi \left( \me^{\mi k t'} \me^{-\mi \phi} x_1 -  \me^{-\mi k t'} \me^{\mi \phi} x_2\right).
\end{align}
Integrating $x_1(t,t')$ from $t'=t$ to $t+\delt$ and making a change of variables results in
\begin{align}
x_1(t,& t+\delt)=x_1(t,t)\me^{-(1/\tau+\mi \Omega)\delt}+ \mi \frac{\rabi}{2} \me^{\mi\phi} \me^{-(1/\tau+\mi \Omega)(t+\delt)}\\ 
& \times \int_0^\delt \dif m \me^{(1/\tau+\mi (\Omega-k))(t+m)} x_3(t,t+m). \label{eq:x1}
\shortintertext{Likewise, for $x_2$ and $x_3$ we get}
x_2(t,& t+\delt)=x_2(t,t)\me^{-(1/\tau-\mi \Omega)\delt} - \mi \frac{\rabi}{2} \me^{-\mi\phi} \me^{-(1/\tau-\mi \Omega)(t+\delt)} \\
&\times \int_0^\delt \dif m \me^{(1/\tau-\mi (\Omega-k))(t+m)} x_3(t,t+m) \label{eq:x2},\\
x_3(t,&t+\delt) = x_3(t,t) \me^{-\frac{2}{\tau}\delt} + \frac{\tau}{2} b(t) (1-\me^{-\frac{2}{\tau}\delt}) \\
&+ \mi \rabi \me^{-\mi \phi} \int_0^\delt \dif m \me^{\frac{2}{\tau}m} \me^{\mi k (t+m)} \me^{-\frac{2}{\tau}\delt} x_1(t,t+m)\\
&-\mi \rabi \me^{\mi \phi} \int_0^\delt \dif m \me^{\frac{2}{\tau}m} \me^{-\mi k (t+m)} \me^{-\frac{2}{\tau}\delt} x_2(t,t+m). \label{eq:x3}
\shortintertext{Substituting \eqref{eq:x1}-\eqref{eq:x2} in \eqref{eq:x3} results in}
x_3(t,&t+\delt) = x_3(t,t) \me^{-\frac{2}{\tau}\delt} + \frac{\tau}{2} b(t) (1-\me^{-\frac{2}{\tau}\delt}) \\
&+ \mi \rabi \me^{-\mi \phi} x_1(t,t) \me^{\mi k t} \frac{\me^{-[\frac{1}{\tau}+\mi(\Omega-k)]\delt}-\me^{-\frac{2}{\tau}\delt}}{\frac{1}{\tau}-\mi(\Omega-k)}\\
&- \mi \rabi \me^{\mi \phi} x_2(t,t) \me^{-\mi k t} \frac{\me^{-[\frac{1}{\tau}-\mi(\Omega-k)]\delt}-\me^{-\frac{2}{\tau}\delt}}{\frac{1}{\tau}+\mi(\Omega-k)}\\
&-\frac{\rabi^2}{2} \int_0^\delt \dif m' x_3(t,t+m') \frac{\me^{-[\frac{1}{\tau}+\mi(\Omega-k)](\delt-m')}-\me^{-\frac{2}{\tau}(\delt-m')}}{\frac{1}{\tau}-\mi(\Omega-k)}\\
&-\frac{\rabi^2}{2} \int_0^\delt \dif m' x_3(t,t+m') \frac{\me^{-[\frac{1}{\tau}-\mi(\Omega-k)](\delt-m')}-\me^{-\frac{2}{\tau}(\delt-m')}}{\frac{1}{\tau}+\mi(\Omega-k)}.
\end{align}
Once we take the Laplace transform of these equations with respect to the $\delt$ variable, the convolution integrals simplify and we are left with
\begin{align}
X_3(s) ={}& \frac{\left(s+\frac{2}{\tau}\right)\left[\left(s+\frac{1}{\tau}\right)^2+(\Omega-k)^2\right]}{\left(s+\frac{2}{\tau}\right)\left[\left(s+\frac{1}{\tau}\right)^2+(\Omega-k)^2\right]+\rabi^2 \left(s+\frac{1}{\tau}\right) }  
\times \\
& \Vast\{x_3(t,t) \frac{1}{s+\frac{2}{\tau}}+\frac{\tau}{2} b(t) \left( \frac{1}{s} - \frac{1}{s+\frac{2}{\tau}} \right)\\
&+\mi \rabi\me^{-\mi\phi} x_1(t,t) \me^{\mi k t}\frac{1}{\left( s+\frac{2}{\tau}\right)\left[s+\frac{1}{\tau}+\mi(\Omega-k)\right]} \\
&- \mi \rabi\me^{\mi\phi} x_2(t,t) \me^{-\mi k t}\frac{1}{\left( s+\frac{2}{\tau}\right)\left[s+\frac{1}{\tau}-\mi(\Omega-k)\right]}
\Vast\}.
\end{align}
Using \eqref{eq:x1}--\eqref{eq:x2} we get
\begin{align}
X_1(s) = x_1(t,t) \frac{1}{s+\frac{1}{\tau}+\mi\Omega} + \mi \frac{\rabi}{2} \me^{\mi \phi} \me^{-\mi k t} X_3(s+\mi k) \frac{1}{s+\frac{1}{\tau}+\mi\Omega}, \\
X_2(s) = x_2(t,t) \frac{1}{s+\frac{1}{\tau}-\mi\Omega} - \mi \frac{\rabi}{2} \me^{-\mi \phi} \me^{\mi k t} X_3(s-\mi k) \frac{1}{s+\frac{1}{\tau}-\mi\Omega}.
\end{align}
These results are the general solution to the Bloch equations expressed in the Laplace domain.


\section{Correlation function calculations} \label{app:B}

\subsection{Single time correlations}
The calculation of $\expect{\sigma_-(t)}$, $\expect{\sigma_+(t)}$, and $\expect{\sigma_z(t)}$ can be made by using the results from the previous section. The inhomogeneous term is $b(t)=-\frac{2}{\tau}$. We assume that the atom is initially in its ground state such that $x_1(0)=x_2(0)=0$, $x_3(0)=-1$. The Laplace transforms of the expectation values are
\begin{align}
\mathcal{L}_t[\me^{\mi k t}\expect{\sigma_-(t)}] ={}& -\mi \frac{\rabi}{2}\me^{\mi\phi} \frac{\left( s+ \frac{2}{\tau}\right)\left(s+\frac{1}{\tau}-\mi(\Omega-k)\right)}{s P(s)}, \label{eq:s1}\quad\\
\mathcal{L}_t[\me^{-\mi k t}\expect{\sigma_+(t)}] ={}& \mi \frac{\rabi}{2}\me^{-\mi\phi} \frac{\left( s+ \frac{2}{\tau}\right)\left(s+\frac{1}{\tau}+\mi(\Omega-k)\right)}{s P(s)},\label{eq:s2}\\
\mathcal{L}_t[\expect{\sigma_z(t)}] ={}& \frac{-\left( s+ \frac{2}{\tau}\right)\left[\left(s+\frac{1}{\tau}\right)^2+(\Omega-k)^2\right]}{s P(s)}, \label{eq:s3}
\end{align}
where
\begin{align}
P(s) = \left(s+\frac{2}{\tau}\right)\left[\left(s+\frac{1}{\tau}\right)^2+(\Omega-k)^2\right]+\rabi^2 \left(s+\frac{1}{\tau}\right). \label{eqn:ps} 
\end{align}
The $t\rightarrow \infty$ limit of these quantities is also of interest. We get
\begin{align}
\lim_{t\rightarrow\infty} \expect{\sigma_-(t)} ={}& \frac{-\frac{\mi}{2} R (1-\mi D)}{1+D^2+\frac{1}{2}R^2} \me^{-\mi k t + \mi \phi},\\
\lim_{t\rightarrow\infty} \expect{\sigma_+(t)} ={}&  \frac{\frac{\mi}{2} R(1+\mi D)}{1+D^2+\frac{1}{2}R^2} \me^{\mi k t - \mi \phi},  \label{eq:spinf}\\
\lim_{t\rightarrow\infty} \frac{\expect{\sigma_z(t)} + 1}{2}={}& \lim_{t\rightarrow\infty} \expect{\sigma_+(t)\sigma_-(t)} =\frac{\frac{1}{4} R^2}{1+D^2+\frac{1}{2}R^2}. \quad \label{eq:szinf}
\end{align}
Here, $D \equiv (\Omega-k)\tau$ stands for the normalized detuning frequency and $R \equiv \rabi \tau$ for the normalized Rabi frequency.

\subsection{Double time correlations of two operators}
In order to calculate $\expect{\sigma_+(t) \sigma_-(t')}$, $\expect{\sigma_+(t) \sigma_+(t')}$ and $\expect{\sigma_+(t) \sigma_z(t')}$ we use initial values at $t'=t$ under steady state conditions when $t\rightarrow \infty$ such that $\expect{\sigma_+(t) \sigma_-(t)}$ is given by \eqref{eq:szinf}, $\expect{\sigma_+(t) \sigma_+(t)} = 0$ and $\expect{\sigma_+(t) \sigma_z(t)} = -\expect{\sigma_+(t)}$ is given by \eqref{eq:spinf}. The inhomogeneous term is $b(t)=-\frac{2}{\tau}\expect{\sigma_+(t)}$. After some algebra we get
\begin{align}
&\mathcal{L}_\delt[\me^{\mi k \delt}\expect{\sigma_+(t) \sigma_-(t+\delt)}] = \frac{\frac{1}{4} R^2}{1+D^2+\frac{1}{2}R^2} \frac{P(s)-\frac{1}{2}\rabi^2\left(s+\frac{2}{\tau}\right)}{s P(s)}, \\
&\mathcal{L}_\delt[\me^{-\mi k \delt}\expect{\sigma_+(t) \sigma_+(t+\delt)}] = \frac{-\frac{1}{4} R^2 \me^{-2\mi\phi+2\mi k t}}{1+D^2+\frac{1}{2}R^2}\\
&\qquad \times \frac{\frac{P(s)-\frac{1}{2}\rabi^2\left(s+\frac{2}{\tau}\right)}{s P(s)} \left[ s+\textstyle\frac{1}{\tau}+\mi(\Omega-k)\right]-1}{s+\frac{1}{\tau}-\mi(\Omega-k)},\\
\end{align}
\begin{align}
&\mathcal{L}_\delt[\expect{\sigma_+(t) \sigma_z(t+\delt)}] = \frac{-\frac{\mi}{2}R \tau \me^{-\mi\phi +\mi k t}}{1+D^2+\frac{1}{2}R^2} \\
&\qquad\times\left\{\frac{P(s)-\frac{1}{2}\rabi^2\left(s+\frac{2}{\tau}\right)}{s P(s)} \left[ s+\textstyle\frac{1}{\tau}+\mi(\Omega-k)\right]-1\right\}.
\end{align}
Note that the Laplace transforms are taken with respect to $\delt$.

\subsection{Double time correlations of three operators}
We multiply input-output equations \eqref{eq:inout} evaluated at time $t'$ by $\sigma_+(t)$ from the left and $\sigma_-(t)$ from the right and take the expectation values to arrive at the double time correlations of three operators. The initial values are given by
\begin{align}
\expect{\sigma_+(t)\sigma_-(t)\sigma_-(t)}={}&\expect{\sigma_+(t)\sigma_+(t)\sigma_-(t)}=0, \\
\expect{\sigma_+(t)\sigma_z(t)\sigma_-(t)}={}&-\expect{\sigma_+(t)\sigma_-(t)}=-\frac{\expect{\sigma_z(t)+1}}{2},
\end{align}
and the inhomogeneous term is
$$
b(t) = -\frac{2}{\tau} \frac{\expect{\sigma_z(t)+1}}{2} .
$$
The expectation value of $\sigma_z(t)$ is at its steady state value given by \eqref{eq:szinf}. If we compare the initial values and the inhomogeneous term to the case of single time correlations, we see that they are exactly the same except for the scaling term $ \expect{\sigma_z(t)+1}/2$ which is given by  \eqref{eq:szinf}. Thus, the results are just rescaled versions of the single time correlation ones and are given by
\begin{align}
\mathcal{L}_\delt[\me^{\mi k \delt}\expect{\sigma_+(t)\sigma_-(t+\delt)\sigma_-(t)}] ={}& \text{\eqref{eq:s1} $\times$ \eqref{eq:szinf}}, \\
\mathcal{L}_\delt[\me^{-\mi k \delt}\expect{\sigma_+(t)\sigma_+(t+\delt)\sigma_-(t)}] ={}&  \text{\eqref{eq:s2} $\times$ \eqref{eq:szinf}},\\
\mathcal{L}_\delt[\expect{\sigma_+(t)\sigma_z(t+\delt)\sigma_-(t)}]={}&  \text{\eqref{eq:s3} $\times$ \eqref{eq:szinf}},
\end{align}
where the Laplace transforms are taken with respect to $\delt$.

\section{Short note on numerics}
The differential equations that we analyzed so far can be transformed into time-independent forms by the substitution 
\begin{align}
\sigmams = \me^{\mi k t} \sigma_-, \quad \sigmaps = \me^{-\mi k t} \sigma_+ \quad \text{and} \quad \sigmazs = \sigma_z,
\end{align}
where $k$ is the frequency of the incoming photons. 
For instance, the single time expectation values of $\sigmams$, $\sigmaps$ and $\sigmazs$ can be written as
\begin{align}
\frac{\dif}{\dif t} 
\begin{pmatrix}
\expect{\sigmams}\\
\expect{\sigmaps}\\
\expect{\sigmazs}
\end{pmatrix}
= \vec{M}
\begin{pmatrix}
\expect{\sigmams}\\
\expect{\sigmaps}\\
\expect{\sigmazs}
\end{pmatrix}
+
\begin{pmatrix}
0\\
0\\
-\frac{2}{\tau}
\end{pmatrix},
\end{align}
where the matrix $\vec{M}$ is given by
\begin{align}
\vec{M} = 
\begin{pmatrix}
-(\frac{1}{\tau} + \mi \Os) & 0 & \mi \frac{\rabi}{2} \me^{\mi \phi} \\ 
0 & -(\frac{1}{\tau} - \mi \Os) & -\mi \frac{\rabi}{2} \me^{-\mi \phi} \\
\mi \rabi \me^{-\mi \phi} & -\mi \rabi \me^{\mi \phi} &  -\frac{2}{\tau}
\end{pmatrix},
\end{align}
and $\Os \equiv \Omega-k$.  Other expectation values have the same form as well. This is a much more convenient formulation for purely numerical studies with which we verified the analytical results reported in the previous appendices.

\bibliography{./fluorescence}

\begin{thebibliography}{34}%
\makeatletter
\providecommand \@ifxundefined [1]{%
 \@ifx{#1\undefined}
}%
\providecommand \@ifnum [1]{%
 \ifnum #1\expandafter \@firstoftwo
 \else \expandafter \@secondoftwo
 \fi
}%
\providecommand \@ifx [1]{%
 \ifx #1\expandafter \@firstoftwo
 \else \expandafter \@secondoftwo
 \fi
}%
\providecommand \natexlab [1]{#1}%
\providecommand \enquote  [1]{``#1''}%
\providecommand \bibnamefont  [1]{#1}%
\providecommand \bibfnamefont [1]{#1}%
\providecommand \citenamefont [1]{#1}%
\providecommand \href@noop [0]{\@secondoftwo}%
\providecommand \href [0]{\begingroup \@sanitize@url \@href}%
\providecommand \@href[1]{\@@startlink{#1}\@@href}%
\providecommand \@@href[1]{\endgroup#1\@@endlink}%
\providecommand \@sanitize@url [0]{\catcode `\\12\catcode `\$12\catcode
  `\&12\catcode `\#12\catcode `\^12\catcode `\_12\catcode `\%12\relax}%
\providecommand \@@startlink[1]{}%
\providecommand \@@endlink[0]{}%
\providecommand \url  [0]{\begingroup\@sanitize@url \@url }%
\providecommand \@url [1]{\endgroup\@href {#1}{\urlprefix }}%
\providecommand \urlprefix  [0]{URL }%
\providecommand \Eprint [0]{\href }%
\providecommand \doibase [0]{http://dx.doi.org/}%
\providecommand \selectlanguage [0]{\@gobble}%
\providecommand \bibinfo  [0]{\@secondoftwo}%
\providecommand \bibfield  [0]{\@secondoftwo}%
\providecommand \translation [1]{[#1]}%
\providecommand \BibitemOpen [0]{}%
\providecommand \bibitemStop [0]{}%
\providecommand \bibitemNoStop [0]{.\EOS\space}%
\providecommand \EOS [0]{\spacefactor3000\relax}%
\providecommand \BibitemShut  [1]{\csname bibitem#1\endcsname}%
\let\auto@bib@innerbib\@empty
\bibitem [{\citenamefont {Mollow}(1969)}]{Mollow1969}%
  \BibitemOpen
  \bibfield  {author} {\bibinfo {author} {\bibfnamefont {B.~R.}\ \bibnamefont
  {Mollow}},\ }\href {\doibase 10.1103/PhysRev.188.1969} {\bibfield  {journal}
  {\bibinfo  {journal} {Phys. Rev.}\ }\textbf {\bibinfo {volume} {188}},\
  \bibinfo {pages} {1969} (\bibinfo {year} {1969})}\BibitemShut {NoStop}%
\bibitem [{\citenamefont {Kimble}\ and\ \citenamefont
  {Mandel}(1976)}]{Kimble1976}%
  \BibitemOpen
  \bibfield  {author} {\bibinfo {author} {\bibfnamefont {H.~J.}\ \bibnamefont
  {Kimble}}\ and\ \bibinfo {author} {\bibfnamefont {L.}~\bibnamefont
  {Mandel}},\ }\href {\doibase 10.1103/PhysRevA.13.2123} {\bibfield  {journal}
  {\bibinfo  {journal} {Phys. Rev. A}\ }\textbf {\bibinfo {volume} {13}},\
  \bibinfo {pages} {2123} (\bibinfo {year} {1976})}\BibitemShut {NoStop}%
\bibitem [{\citenamefont {O'Brien}\ \emph {et~al.}(2009)\citenamefont
  {O'Brien}, \citenamefont {Furusawa},\ and\ \citenamefont
  {Vuckovic}}]{OBrien2009}%
  \BibitemOpen
  \bibfield  {author} {\bibinfo {author} {\bibfnamefont {J.~L.}\ \bibnamefont
  {O'Brien}}, \bibinfo {author} {\bibfnamefont {A.}~\bibnamefont {Furusawa}}, \
  and\ \bibinfo {author} {\bibfnamefont {J.}~\bibnamefont {Vuckovic}},\ }\href
  {http://dx.doi.org/10.1038/nphoton.2009.229} {\bibfield  {journal} {\bibinfo
  {journal} {Nat Photon}\ }\textbf {\bibinfo {volume} {3}},\ \bibinfo {pages}
  {687} (\bibinfo {year} {2009})}\BibitemShut {NoStop}%
\bibitem [{\citenamefont {Benson}(2011)}]{Benson2011}%
  \BibitemOpen
  \bibfield  {author} {\bibinfo {author} {\bibfnamefont {O.}~\bibnamefont
  {Benson}},\ }\href {http://dx.doi.org/10.1038/nature10610} {\bibfield
  {journal} {\bibinfo  {journal} {Nature}\ }\textbf {\bibinfo {volume} {480}},\
  \bibinfo {pages} {193} (\bibinfo {year} {2011})}\BibitemShut {NoStop}%
\bibitem [{\citenamefont {You}\ and\ \citenamefont {Nori}(2011)}]{You2011}%
  \BibitemOpen
  \bibfield  {author} {\bibinfo {author} {\bibfnamefont {J.~Q.}\ \bibnamefont
  {You}}\ and\ \bibinfo {author} {\bibfnamefont {F.}~\bibnamefont {Nori}},\
  }\href {http://dx.doi.org/10.1038/nature10122} {\bibfield  {journal}
  {\bibinfo  {journal} {Nature}\ }\textbf {\bibinfo {volume} {474}},\ \bibinfo
  {pages} {589} (\bibinfo {year} {2011})}\BibitemShut {NoStop}%
\bibitem [{\citenamefont {Astafiev}\ \emph {et~al.}(2010)\citenamefont
  {Astafiev}, \citenamefont {Zagoskin}, \citenamefont {Abdumalikov},
  \citenamefont {Pashkin}, \citenamefont {Yamamoto}, \citenamefont {Inomata},
  \citenamefont {Nakamura},\ and\ \citenamefont {Tsai}}]{Astafiev2010}%
  \BibitemOpen
  \bibfield  {author} {\bibinfo {author} {\bibfnamefont {O.}~\bibnamefont
  {Astafiev}}, \bibinfo {author} {\bibfnamefont {A.~M.}\ \bibnamefont
  {Zagoskin}}, \bibinfo {author} {\bibfnamefont {J.}~\bibnamefont
  {Abdumalikov}, \bibfnamefont {A.~A.}}, \bibinfo {author} {\bibfnamefont
  {Y.~A.}\ \bibnamefont {Pashkin}}, \bibinfo {author} {\bibfnamefont
  {T.}~\bibnamefont {Yamamoto}}, \bibinfo {author} {\bibfnamefont
  {K.}~\bibnamefont {Inomata}}, \bibinfo {author} {\bibfnamefont
  {Y.}~\bibnamefont {Nakamura}}, \ and\ \bibinfo {author} {\bibfnamefont
  {J.~S.}\ \bibnamefont {Tsai}},\ }\href {\doibase 10.1126/science.1181918}
  {\bibfield  {journal} {\bibinfo  {journal} {Science}\ }\textbf {\bibinfo
  {volume} {327}},\ \bibinfo {pages} {840} (\bibinfo {year}
  {2010})}\BibitemShut {NoStop}%
\bibitem [{\citenamefont {Bozyigit}\ \emph {et~al.}(2011)\citenamefont
  {Bozyigit}, \citenamefont {Lang}, \citenamefont {Steffen}, \citenamefont
  {Fink}, \citenamefont {Eichler}, \citenamefont {Baur}, \citenamefont
  {Bianchetti}, \citenamefont {Leek}, \citenamefont {Filipp}, \citenamefont
  {da~Silva}, \citenamefont {Blais},\ and\ \citenamefont
  {Wallraff}}]{Bozyigit2011}%
  \BibitemOpen
  \bibfield  {author} {\bibinfo {author} {\bibfnamefont {D.}~\bibnamefont
  {Bozyigit}}, \bibinfo {author} {\bibfnamefont {C.}~\bibnamefont {Lang}},
  \bibinfo {author} {\bibfnamefont {L.}~\bibnamefont {Steffen}}, \bibinfo
  {author} {\bibfnamefont {J.~M.}\ \bibnamefont {Fink}}, \bibinfo {author}
  {\bibfnamefont {C.}~\bibnamefont {Eichler}}, \bibinfo {author} {\bibfnamefont
  {M.}~\bibnamefont {Baur}}, \bibinfo {author} {\bibfnamefont {R.}~\bibnamefont
  {Bianchetti}}, \bibinfo {author} {\bibfnamefont {P.~J.}\ \bibnamefont
  {Leek}}, \bibinfo {author} {\bibfnamefont {S.}~\bibnamefont {Filipp}},
  \bibinfo {author} {\bibfnamefont {M.~P.}\ \bibnamefont {da~Silva}}, \bibinfo
  {author} {\bibfnamefont {A.}~\bibnamefont {Blais}}, \ and\ \bibinfo {author}
  {\bibfnamefont {A.}~\bibnamefont {Wallraff}},\ }\href
  {http://dx.doi.org/10.1038/nphys1845} {\bibfield  {journal} {\bibinfo
  {journal} {Nat Phys}\ }\textbf {\bibinfo {volume} {7}},\ \bibinfo {pages}
  {154} (\bibinfo {year} {2011})}\BibitemShut {NoStop}%
\bibitem [{\citenamefont {da~Silva}\ \emph {et~al.}(2010)\citenamefont
  {da~Silva}, \citenamefont {Bozyigit}, \citenamefont {Wallraff},\ and\
  \citenamefont {Blais}}]{Silva2010}%
  \BibitemOpen
  \bibfield  {author} {\bibinfo {author} {\bibfnamefont {M.~P.}\ \bibnamefont
  {da~Silva}}, \bibinfo {author} {\bibfnamefont {D.}~\bibnamefont {Bozyigit}},
  \bibinfo {author} {\bibfnamefont {A.}~\bibnamefont {Wallraff}}, \ and\
  \bibinfo {author} {\bibfnamefont {A.}~\bibnamefont {Blais}},\ }\href
  {\doibase 10.1103/PhysRevA.82.043804} {\bibfield  {journal} {\bibinfo
  {journal} {Phys. Rev. A}\ }\textbf {\bibinfo {volume} {82}},\ \bibinfo
  {pages} {043804} (\bibinfo {year} {2010})}\BibitemShut {NoStop}%
\bibitem [{\citenamefont {Abdumalikov}\ \emph {et~al.}(2011)\citenamefont
  {Abdumalikov}, \citenamefont {Astafiev}, \citenamefont {Pashkin},
  \citenamefont {Nakamura},\ and\ \citenamefont {Tsai}}]{Abdumalikov2011}%
  \BibitemOpen
  \bibfield  {author} {\bibinfo {author} {\bibfnamefont {A.~A.}\ \bibnamefont
  {Abdumalikov}}, \bibinfo {author} {\bibfnamefont {O.~V.}\ \bibnamefont
  {Astafiev}}, \bibinfo {author} {\bibfnamefont {Y.~A.}\ \bibnamefont
  {Pashkin}}, \bibinfo {author} {\bibfnamefont {Y.}~\bibnamefont {Nakamura}}, \
  and\ \bibinfo {author} {\bibfnamefont {J.~S.}\ \bibnamefont {Tsai}},\ }\href
  {\doibase 10.1103/PhysRevLett.107.043604} {\bibfield  {journal} {\bibinfo
  {journal} {Phys. Rev. Lett.}\ }\textbf {\bibinfo {volume} {107}},\ \bibinfo
  {pages} {043604} (\bibinfo {year} {2011})}\BibitemShut {NoStop}%
\bibitem [{\citenamefont {Abdumalikov}\ \emph {et~al.}(2010)\citenamefont
  {Abdumalikov}, \citenamefont {Astafiev}, \citenamefont {Zagoskin},
  \citenamefont {Pashkin}, \citenamefont {Nakamura},\ and\ \citenamefont
  {Tsai}}]{Abdumalikov2010}%
  \BibitemOpen
  \bibfield  {author} {\bibinfo {author} {\bibfnamefont {A.~A.}\ \bibnamefont
  {Abdumalikov}}, \bibinfo {author} {\bibfnamefont {O.}~\bibnamefont
  {Astafiev}}, \bibinfo {author} {\bibfnamefont {A.~M.}\ \bibnamefont
  {Zagoskin}}, \bibinfo {author} {\bibfnamefont {Y.~A.}\ \bibnamefont
  {Pashkin}}, \bibinfo {author} {\bibfnamefont {Y.}~\bibnamefont {Nakamura}}, \
  and\ \bibinfo {author} {\bibfnamefont {J.~S.}\ \bibnamefont {Tsai}},\ }\href
  {\doibase 10.1103/PhysRevLett.104.193601} {\bibfield  {journal} {\bibinfo
  {journal} {Phys. Rev. Lett.}\ }\textbf {\bibinfo {volume} {104}},\ \bibinfo
  {pages} {193601} (\bibinfo {year} {2010})}\BibitemShut {NoStop}%
\bibitem [{\citenamefont {Hoi}\ \emph {et~al.}(2011)\citenamefont {Hoi},
  \citenamefont {Wilson}, \citenamefont {Johansson}, \citenamefont {Palomaki},
  \citenamefont {Peropadre},\ and\ \citenamefont {Delsing}}]{Hoi2011}%
  \BibitemOpen
  \bibfield  {author} {\bibinfo {author} {\bibfnamefont {I.-C.}\ \bibnamefont
  {Hoi}}, \bibinfo {author} {\bibfnamefont {C.~M.}\ \bibnamefont {Wilson}},
  \bibinfo {author} {\bibfnamefont {G.}~\bibnamefont {Johansson}}, \bibinfo
  {author} {\bibfnamefont {T.}~\bibnamefont {Palomaki}}, \bibinfo {author}
  {\bibfnamefont {B.}~\bibnamefont {Peropadre}}, \ and\ \bibinfo {author}
  {\bibfnamefont {P.}~\bibnamefont {Delsing}},\ }\href {\doibase
  10.1103/PhysRevLett.107.073601} {\bibfield  {journal} {\bibinfo  {journal}
  {Phys. Rev. Lett.}\ }\textbf {\bibinfo {volume} {107}},\ \bibinfo {pages}
  {073601} (\bibinfo {year} {2011})}\BibitemShut {NoStop}%
\bibitem [{\citenamefont {Florescu}\ and\ \citenamefont
  {John}(2004)}]{Florescu2004}%
  \BibitemOpen
  \bibfield  {author} {\bibinfo {author} {\bibfnamefont {M.}~\bibnamefont
  {Florescu}}\ and\ \bibinfo {author} {\bibfnamefont {S.}~\bibnamefont
  {John}},\ }\href {\doibase 10.1103/PhysRevA.69.053810} {\bibfield  {journal}
  {\bibinfo  {journal} {Phys. Rev. A}\ }\textbf {\bibinfo {volume} {69}},\
  \bibinfo {pages} {053810} (\bibinfo {year} {2004})}\BibitemShut {NoStop}%
\bibitem [{\citenamefont {Davan\c{c}o}\ and\ \citenamefont
  {Srinivasan}(2009)}]{Davanco2009}%
  \BibitemOpen
  \bibfield  {author} {\bibinfo {author} {\bibfnamefont {M.}~\bibnamefont
  {Davan\c{c}o}}\ and\ \bibinfo {author} {\bibfnamefont {K.}~\bibnamefont
  {Srinivasan}},\ }\href {http://ol.osa.org/abstract.cfm?URI=ol-34-16-2542}
  {\bibfield  {journal} {\bibinfo  {journal} {Opt. Lett.}\ }\textbf {\bibinfo
  {volume} {34}},\ \bibinfo {pages} {2542} (\bibinfo {year}
  {2009})}\BibitemShut {NoStop}%
\bibitem [{\citenamefont {Davan\c{c}o}\ \emph {et~al.}(2011)\citenamefont
  {Davan\c{c}o}, \citenamefont {Rakher}, \citenamefont {Wegscheider},
  \citenamefont {Schuh}, \citenamefont {Badolato},\ and\ \citenamefont
  {Srinivasan}}]{Davanco2011a}%
  \BibitemOpen
  \bibfield  {author} {\bibinfo {author} {\bibfnamefont {M.}~\bibnamefont
  {Davan\c{c}o}}, \bibinfo {author} {\bibfnamefont {M.~T.}\ \bibnamefont
  {Rakher}}, \bibinfo {author} {\bibfnamefont {W.}~\bibnamefont {Wegscheider}},
  \bibinfo {author} {\bibfnamefont {D.}~\bibnamefont {Schuh}}, \bibinfo
  {author} {\bibfnamefont {A.}~\bibnamefont {Badolato}}, \ and\ \bibinfo
  {author} {\bibfnamefont {K.}~\bibnamefont {Srinivasan}},\ }\href {\doibase
  10.1063/1.3617472} {\bibfield  {journal} {\bibinfo  {journal} {Appl. Phys.
  Lett.}\ }\textbf {\bibinfo {volume} {99}},\ \bibinfo {pages} {121101}
  (\bibinfo {year} {2011})}\BibitemShut {NoStop}%
\bibitem [{\citenamefont {Mandel}\ and\ \citenamefont
  {Wolf}(1995)}]{Mandel1995}%
  \BibitemOpen
  \bibfield  {author} {\bibinfo {author} {\bibfnamefont {L.}~\bibnamefont
  {Mandel}}\ and\ \bibinfo {author} {\bibfnamefont {E.}~\bibnamefont {Wolf}},\
  }\href@noop {} {\emph {\bibinfo {title} {Optical coherence and quantum
  optics}}}\ (\bibinfo  {publisher} {Cambridge University Press},\ \bibinfo
  {year} {1995})\ \bibinfo {note} {sec. 15.6}\BibitemShut {NoStop}%
\bibitem [{\citenamefont {Carmichael}(2002)}]{Carmichael2002}%
  \BibitemOpen
  \bibfield  {author} {\bibinfo {author} {\bibfnamefont {H.}~\bibnamefont
  {Carmichael}},\ }\href@noop {} {\emph {\bibinfo {title} {Statistical Methods
  in Quantum Optics 1: Master Equations and Fokker-Planck Equations}}}\
  (\bibinfo  {publisher} {Springer},\ \bibinfo {year} {2002})\ \bibinfo {note}
  {sec. 2.3}\BibitemShut {NoStop}%
\bibitem [{\citenamefont {Scully}\ and\ \citenamefont
  {Zubairy}(1997)}]{Scully2008}%
  \BibitemOpen
  \bibfield  {author} {\bibinfo {author} {\bibfnamefont {M.~O.}\ \bibnamefont
  {Scully}}\ and\ \bibinfo {author} {\bibfnamefont {M.~S.}\ \bibnamefont
  {Zubairy}},\ }\href@noop {} {\emph {\bibinfo {title} {Quantum optics}}}\
  (\bibinfo  {publisher} {Cambridge University Press},\ \bibinfo {year}
  {1997})\ \bibinfo {note} {chap. 10}\BibitemShut {NoStop}%
\bibitem [{\citenamefont {Walls}\ and\ \citenamefont
  {Milburn}(2008)}]{Walls2008}%
  \BibitemOpen
  \bibfield  {author} {\bibinfo {author} {\bibfnamefont {D.}~\bibnamefont
  {Walls}}\ and\ \bibinfo {author} {\bibfnamefont {G.~J.}\ \bibnamefont
  {Milburn}},\ }\href {\doibase 10.1007/978-3-540-28574-8} {\emph {\bibinfo
  {title} {Quantum Optics}}},\ \bibinfo {edition} {2nd}\ ed.\ (\bibinfo
  {publisher} {Springer},\ \bibinfo {year} {2008})\ \bibinfo {note} {sec.
  10.5}\BibitemShut {NoStop}%
\bibitem [{\citenamefont {Hennrich}\ \emph {et~al.}(2005)\citenamefont
  {Hennrich}, \citenamefont {Kuhn},\ and\ \citenamefont
  {Rempe}}]{Hennrich2005}%
  \BibitemOpen
  \bibfield  {author} {\bibinfo {author} {\bibfnamefont {M.}~\bibnamefont
  {Hennrich}}, \bibinfo {author} {\bibfnamefont {A.}~\bibnamefont {Kuhn}}, \
  and\ \bibinfo {author} {\bibfnamefont {G.}~\bibnamefont {Rempe}},\ }\href
  {\doibase 10.1103/PhysRevLett.94.053604} {\bibfield  {journal} {\bibinfo
  {journal} {Phys. Rev. Lett.}\ }\textbf {\bibinfo {volume} {94}},\ \bibinfo
  {pages} {053604} (\bibinfo {year} {2005})}\BibitemShut {NoStop}%
\bibitem [{\citenamefont {Le~Kien}\ and\ \citenamefont
  {Hakuta}(2008)}]{Kien2008}%
  \BibitemOpen
  \bibfield  {author} {\bibinfo {author} {\bibfnamefont {F.}~\bibnamefont
  {Le~Kien}}\ and\ \bibinfo {author} {\bibfnamefont {K.}~\bibnamefont
  {Hakuta}},\ }\href {\doibase 10.1103/PhysRevA.77.033826} {\bibfield
  {journal} {\bibinfo  {journal} {Phys. Rev. A}\ }\textbf {\bibinfo {volume}
  {77}},\ \bibinfo {pages} {033826} (\bibinfo {year} {2008})}\BibitemShut
  {NoStop}%
\bibitem [{\citenamefont {Jin}\ \emph {et~al.}(2011)\citenamefont {Jin},
  \citenamefont {Evers},\ and\ \citenamefont {Macovei}}]{Jin2011}%
  \BibitemOpen
  \bibfield  {author} {\bibinfo {author} {\bibfnamefont {L.}~\bibnamefont
  {Jin}}, \bibinfo {author} {\bibfnamefont {J.}~\bibnamefont {Evers}}, \ and\
  \bibinfo {author} {\bibfnamefont {M.}~\bibnamefont {Macovei}},\ }\href
  {\doibase 10.1103/PhysRevA.84.043812} {\bibfield  {journal} {\bibinfo
  {journal} {Phys. Rev. A}\ }\textbf {\bibinfo {volume} {84}},\ \bibinfo
  {pages} {043812} (\bibinfo {year} {2011})}\BibitemShut {NoStop}%
\bibitem [{\citenamefont {Shen}\ and\ \citenamefont
  {Fan}(2007{\natexlab{a}})}]{Shen2007a}%
  \BibitemOpen
  \bibfield  {author} {\bibinfo {author} {\bibfnamefont {J.~T.}\ \bibnamefont
  {Shen}}\ and\ \bibinfo {author} {\bibfnamefont {S.}~\bibnamefont {Fan}},\
  }\href {\doibase 10.1103/PhysRevLett.98.153003} {\bibfield  {journal}
  {\bibinfo  {journal} {Phys. Rev. Lett.}\ }\textbf {\bibinfo {volume} {98}},\
  \bibinfo {pages} {153003} (\bibinfo {year} {2007}{\natexlab{a}})}\BibitemShut
  {NoStop}%
\bibitem [{\citenamefont {Shen}\ and\ \citenamefont
  {Fan}(2007{\natexlab{b}})}]{Shen2007}%
  \BibitemOpen
  \bibfield  {author} {\bibinfo {author} {\bibfnamefont {J.~T.}\ \bibnamefont
  {Shen}}\ and\ \bibinfo {author} {\bibfnamefont {S.}~\bibnamefont {Fan}},\
  }\href {\doibase 10.1103/PhysRevA.76.062709} {\bibfield  {journal} {\bibinfo
  {journal} {Physical Review A}\ }\textbf {\bibinfo {volume} {76}},\ \bibinfo
  {pages} {062709} (\bibinfo {year} {2007}{\natexlab{b}})}\BibitemShut
  {NoStop}%
\bibitem [{\citenamefont {Gardiner}\ and\ \citenamefont
  {Collett}(1985)}]{Gardiner1985}%
  \BibitemOpen
  \bibfield  {author} {\bibinfo {author} {\bibfnamefont {C.~W.}\ \bibnamefont
  {Gardiner}}\ and\ \bibinfo {author} {\bibfnamefont {M.~J.}\ \bibnamefont
  {Collett}},\ }\href {\doibase 10.1103/PhysRevA.31.3761} {\bibfield  {journal}
  {\bibinfo  {journal} {Phys. Rev. A}\ }\textbf {\bibinfo {volume} {31}},\
  \bibinfo {pages} {3761} (\bibinfo {year} {1985})}\BibitemShut {NoStop}%
\bibitem [{\citenamefont {Fan}\ \emph {et~al.}(2010)\citenamefont {Fan},
  \citenamefont {Kocaba\c{s}},\ and\ \citenamefont {Shen}}]{Fan2010}%
  \BibitemOpen
  \bibfield  {author} {\bibinfo {author} {\bibfnamefont {S.}~\bibnamefont
  {Fan}}, \bibinfo {author} {\bibfnamefont {{\c{S}}.~E.}\ \bibnamefont
  {Kocaba\c{s}}}, \ and\ \bibinfo {author} {\bibfnamefont {J.-T.}\ \bibnamefont
  {Shen}},\ }\href {\doibase 10.1103/PhysRevA.82.063821} {\bibfield  {journal}
  {\bibinfo  {journal} {Phys. Rev. A}\ }\textbf {\bibinfo {volume} {82}},\
  \bibinfo {pages} {063821} (\bibinfo {year} {2010})}\BibitemShut {NoStop}%
\bibitem [{\citenamefont {Rephaeli}\ \emph {et~al.}(2011)\citenamefont
  {Rephaeli}, \citenamefont {Kocaba\c{s}},\ and\ \citenamefont
  {Fan}}]{Rephaeli2011}%
  \BibitemOpen
  \bibfield  {author} {\bibinfo {author} {\bibfnamefont {E.}~\bibnamefont
  {Rephaeli}}, \bibinfo {author} {\bibfnamefont {{\c{S}}.~E.}\ \bibnamefont
  {Kocaba\c{s}}}, \ and\ \bibinfo {author} {\bibfnamefont {S.}~\bibnamefont
  {Fan}},\ }\href {\doibase 10.1103/PhysRevA.84.063832} {\bibfield  {journal}
  {\bibinfo  {journal} {Phys. Rev. A}\ }\textbf {\bibinfo {volume} {84}},\
  \bibinfo {pages} {063832} (\bibinfo {year} {2011})}\BibitemShut {NoStop}%
\bibitem [{\citenamefont {Barnett}\ and\ \citenamefont
  {Radmore}(1997)}]{Barnett1997}%
  \BibitemOpen
  \bibfield  {author} {\bibinfo {author} {\bibfnamefont {S.~M.}\ \bibnamefont
  {Barnett}}\ and\ \bibinfo {author} {\bibfnamefont {P.~M.}\ \bibnamefont
  {Radmore}},\ }\href@noop {} {\emph {\bibinfo {title} {Methods in theoretical
  quantum optics}}}\ (\bibinfo  {publisher} {Oxford University Press},\
  \bibinfo {year} {1997})\ \bibinfo {note} {sec. 3.6}\BibitemShut {NoStop}%
\bibitem [{Sup()}]{Supplement}%
  \BibitemOpen
  \href@noop {} {}\bibinfo {note} {See Supplemental Material at [URL will be
  inserted by publisher] for the Mathematica source code used in the
  derivations.}\BibitemShut {Stop}%
\bibitem [{\citenamefont {Chang}\ \emph {et~al.}(2007)\citenamefont {Chang},
  \citenamefont {S{\o}rensen}, \citenamefont {Demler},\ and\ \citenamefont
  {Lukin}}]{Chang2007}%
  \BibitemOpen
  \bibfield  {author} {\bibinfo {author} {\bibfnamefont {D.}~\bibnamefont
  {Chang}}, \bibinfo {author} {\bibfnamefont {A.}~\bibnamefont {S{\o}rensen}},
  \bibinfo {author} {\bibfnamefont {E.}~\bibnamefont {Demler}}, \ and\ \bibinfo
  {author} {\bibfnamefont {M.}~\bibnamefont {Lukin}},\ }\href {\doibase
  10.1038/nphys708} {\bibfield  {journal} {\bibinfo  {journal} {Nature
  Physics}\ }\textbf {\bibinfo {volume} {3}},\ \bibinfo {pages} {807} (\bibinfo
  {year} {2007})}\BibitemShut {NoStop}%
\bibitem [{\citenamefont {Zheng}\ \emph {et~al.}(2010)\citenamefont {Zheng},
  \citenamefont {Gauthier},\ and\ \citenamefont {Baranger}}]{Zheng2010}%
  \BibitemOpen
  \bibfield  {author} {\bibinfo {author} {\bibfnamefont {H.}~\bibnamefont
  {Zheng}}, \bibinfo {author} {\bibfnamefont {D.~J.}\ \bibnamefont {Gauthier}},
  \ and\ \bibinfo {author} {\bibfnamefont {H.~U.}\ \bibnamefont {Baranger}},\
  }\href {\doibase 10.1103/PhysRevA.82.063816} {\bibfield  {journal} {\bibinfo
  {journal} {Phys. Rev. A}\ }\textbf {\bibinfo {volume} {82}},\ \bibinfo
  {pages} {063816} (\bibinfo {year} {2010})}\BibitemShut {NoStop}%
\bibitem [{\citenamefont {Hoi}\ \emph {et~al.}(2012)\citenamefont {Hoi},
  \citenamefont {Palomaki}, \citenamefont {Johansson}, \citenamefont
  {Lindkvist}, \citenamefont {Delsing},\ and\ \citenamefont
  {Wilson}}]{Hoi2012}%
  \BibitemOpen
  \bibfield  {author} {\bibinfo {author} {\bibfnamefont {I.-C.}\ \bibnamefont
  {Hoi}}, \bibinfo {author} {\bibfnamefont {T.}~\bibnamefont {Palomaki}},
  \bibinfo {author} {\bibfnamefont {G.}~\bibnamefont {Johansson}}, \bibinfo
  {author} {\bibfnamefont {J.}~\bibnamefont {Lindkvist}}, \bibinfo {author}
  {\bibfnamefont {P.}~\bibnamefont {Delsing}}, \ and\ \bibinfo {author}
  {\bibfnamefont {C.~M.}\ \bibnamefont {Wilson}},\ }\href@noop {} {\  (\bibinfo
  {year} {2012})},\ \Eprint {http://arxiv.org/abs/1201.2269} {1201.2269}
  \BibitemShut {NoStop}%
\bibitem [{\citenamefont {Schuster}\ \emph {et~al.}(2005)\citenamefont
  {Schuster}, \citenamefont {Wallraff}, \citenamefont {Blais}, \citenamefont
  {Frunzio}, \citenamefont {Huang}, \citenamefont {Majer}, \citenamefont
  {Girvin},\ and\ \citenamefont {Schoelkopf}}]{Schuster2005}%
  \BibitemOpen
  \bibfield  {author} {\bibinfo {author} {\bibfnamefont {D.~I.}\ \bibnamefont
  {Schuster}}, \bibinfo {author} {\bibfnamefont {A.}~\bibnamefont {Wallraff}},
  \bibinfo {author} {\bibfnamefont {A.}~\bibnamefont {Blais}}, \bibinfo
  {author} {\bibfnamefont {L.}~\bibnamefont {Frunzio}}, \bibinfo {author}
  {\bibfnamefont {R.-S.}\ \bibnamefont {Huang}}, \bibinfo {author}
  {\bibfnamefont {J.}~\bibnamefont {Majer}}, \bibinfo {author} {\bibfnamefont
  {S.~M.}\ \bibnamefont {Girvin}}, \ and\ \bibinfo {author} {\bibfnamefont
  {R.~J.}\ \bibnamefont {Schoelkopf}},\ }\href {\doibase
  10.1103/PhysRevLett.94.123602} {\bibfield  {journal} {\bibinfo  {journal}
  {Phys. Rev. Lett.}\ }\textbf {\bibinfo {volume} {94}},\ \bibinfo {pages}
  {123602} (\bibinfo {year} {2005})}\BibitemShut {NoStop}%
\bibitem [{\citenamefont {Fragner}\ \emph {et~al.}(2008)\citenamefont
  {Fragner}, \citenamefont {Göppl}, \citenamefont {Fink}, \citenamefont {Baur},
  \citenamefont {Bianchetti}, \citenamefont {Leek}, \citenamefont {Blais},\
  and\ \citenamefont {Wallraff}}]{Fragner2008}%
  \BibitemOpen
  \bibfield  {author} {\bibinfo {author} {\bibfnamefont {A.}~\bibnamefont
  {Fragner}}, \bibinfo {author} {\bibfnamefont {M.}~\bibnamefont {Göppl}},
  \bibinfo {author} {\bibfnamefont {J.~M.}\ \bibnamefont {Fink}}, \bibinfo
  {author} {\bibfnamefont {M.}~\bibnamefont {Baur}}, \bibinfo {author}
  {\bibfnamefont {R.}~\bibnamefont {Bianchetti}}, \bibinfo {author}
  {\bibfnamefont {P.~J.}\ \bibnamefont {Leek}}, \bibinfo {author}
  {\bibfnamefont {A.}~\bibnamefont {Blais}}, \ and\ \bibinfo {author}
  {\bibfnamefont {A.}~\bibnamefont {Wallraff}},\ }\href {\doibase
  10.1126/science.1164482} {\bibfield  {journal} {\bibinfo  {journal}
  {Science}\ }\textbf {\bibinfo {volume} {322}},\ \bibinfo {pages} {1357}
  (\bibinfo {year} {2008})}\BibitemShut {NoStop}%
\bibitem [{\citenamefont {Koch}\ \emph {et~al.}(2011)\citenamefont {Koch},
  \citenamefont {Sames}, \citenamefont {Balbach}, \citenamefont {Chibani},
  \citenamefont {Kubanek}, \citenamefont {Murr}, \citenamefont {Wilk},\ and\
  \citenamefont {Rempe}}]{Koch2011}%
  \BibitemOpen
  \bibfield  {author} {\bibinfo {author} {\bibfnamefont {M.}~\bibnamefont
  {Koch}}, \bibinfo {author} {\bibfnamefont {C.}~\bibnamefont {Sames}},
  \bibinfo {author} {\bibfnamefont {M.}~\bibnamefont {Balbach}}, \bibinfo
  {author} {\bibfnamefont {H.}~\bibnamefont {Chibani}}, \bibinfo {author}
  {\bibfnamefont {A.}~\bibnamefont {Kubanek}}, \bibinfo {author} {\bibfnamefont
  {K.}~\bibnamefont {Murr}}, \bibinfo {author} {\bibfnamefont {T.}~\bibnamefont
  {Wilk}}, \ and\ \bibinfo {author} {\bibfnamefont {G.}~\bibnamefont {Rempe}},\
  }\href {\doibase 10.1103/PhysRevLett.107.023601} {\bibfield  {journal}
  {\bibinfo  {journal} {Phys. Rev. Lett.}\ }\textbf {\bibinfo {volume} {107}},\
  \bibinfo {pages} {023601} (\bibinfo {year} {2011})}\BibitemShut {NoStop}%
\end{thebibliography}%

\end{document}